\documentclass[aps,prb,twocolumn,groupedaddress,citeautoscript,nopacs]{revtex4}

\usepackage{graphicx}
\usepackage{color}
\usepackage{amsfonts}
\setcitestyle{super}

\usepackage{bibunits}

\begin{document}

\begin{bibunit}[naturemag]

\title{Emergent quantum confinement at topological insulator surfaces} 

\author{M.~S.~Bahramy,$^{1,*}$ P.~D.~C.~King,$^{2,*,\dag}$ A.~de~la~Torre,$^2$ J.~Chang,$^3$ M.~Shi,$^3$ L.~Patthey,$^3$ G.~Balakrishnan,$^4$ Ph.~Hofmann,$^5$ R.~Arita,$^{1,6}$   N.~Nagaosa,$^{1,6,7}$ and F.~Baumberger$^2$}

\affiliation{$^1$ Correlated Electron Research Group (CERG), RIKEN-ASI, Wako, Saitama 351-0198, Japan}

\affiliation{$^2$ SUPA, School of Physics and Astronomy, University of St. Andrews, St. Andrews, Fife KY16 9SS, United Kingdom}

\affiliation{$^3$ Swiss Light Source, Paul Scherrer Institut, CH-5232 Villigen PSI, Switzerland}

\affiliation{$^4$ Department of Physics, University of Warwick, Coventry CV4 7AL, United Kingdom}

\affiliation{$^5$ Department of Physics and Astronomy, Interdisciplinary Nanoscience Center, Aarhus University, 8000 Aarhus C, Denmark}

\affiliation{$^6$ Department of Applied Physics, University of Tokyo, Tokyo 113-8656, Japan}

\affiliation{$^7$ Cross-Correlated Materials Research Group (CMRG), RIKEN-ASI, Wako, Saitama 351-0198, Japan}

\affiliation{$^*$ These authors contributed equally to this work}

\affiliation{$^\dag$ To whom correspondence should be addressed; E-mail: philip.king@st-andrews.ac.uk}
\date{\today}

\maketitle

\begin{figure}
\begin{center}
\includegraphics[width=0.5\textwidth]{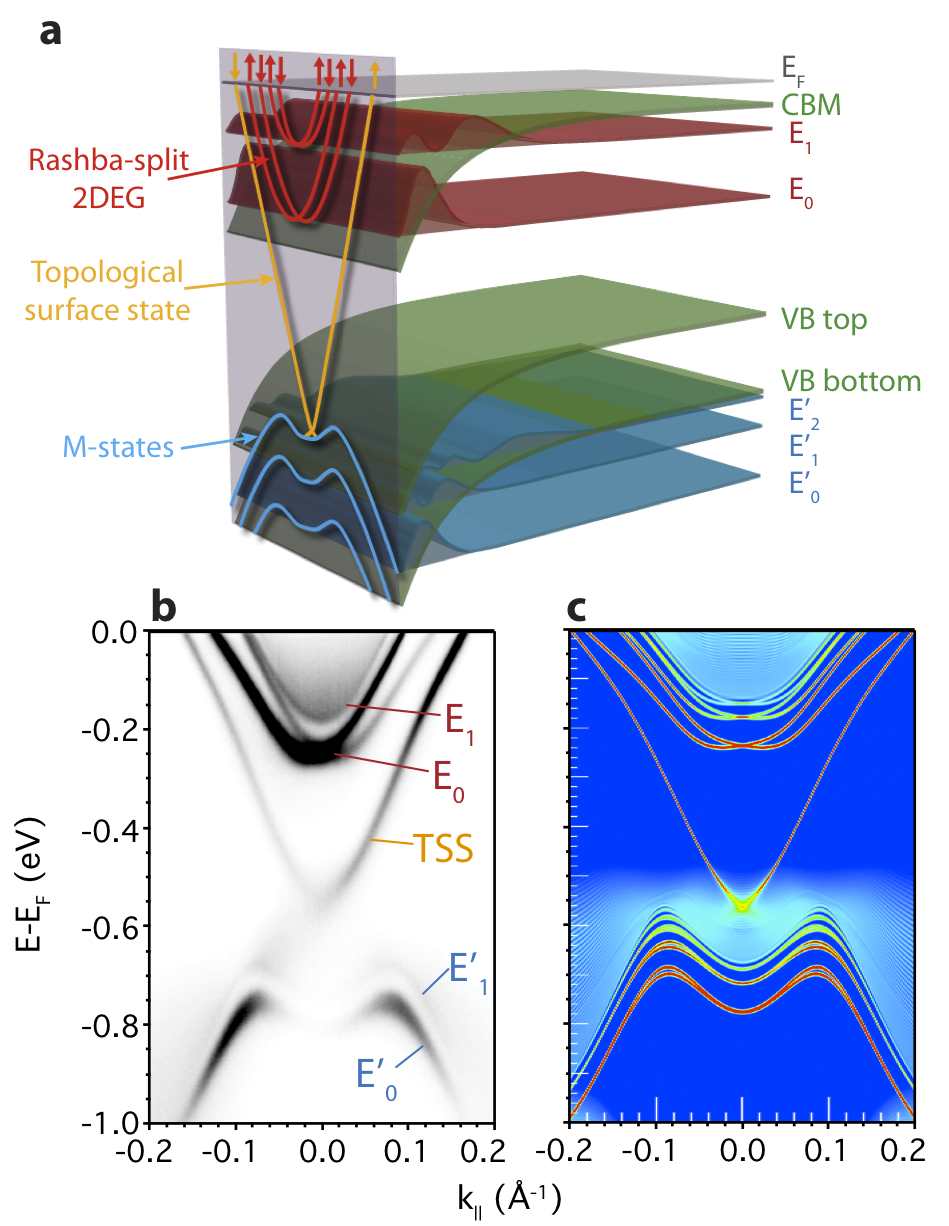}
\caption{ \label{f:overview} {\bf Band-bending-induced surface electronic structure of Bi$_2$Se$_3$.} (a) A near-surface electrostatic potential variation forms a surface quantum well. This shifts the Dirac point of the topological surface state (TSS) further below the Fermi level and causes the conduction band states to be restructured into ladders of Rashba-split two-dimensional subbands, with envelope wave functions (dark red) peaked close to the surface. The finite valence band width provides simultaneous confinement of the valence band states, causing these to also become quantized into ladders of M-shaped states. These features can all be seen in ARPES measurements of the surface electronic structure (b) and are well described by the surface projection ({\it i.e.} real-space projection onto the first 3 QL) of an {\it ab-initio}-derived tight-binding model (c) of the ideal Bi$_2$Se$_3$ bulk structure subject only to a perturbing electrostatic potential close to the surface.}
\end{center}
\end{figure}
{\bf Bismuth-chalchogenides are model examples of three-dimensional topological insulators.~\cite{Hasan:Rev.Mod.Phys.:82(2010)3045--3067} Their ideal bulk-truncated surface hosts a single spin-helical surface state,~\cite{Zhang:NaturePhys.:5(2009)438--442,Xia:NaturePhys.:5(2009)398--402} which is the simplest possible surface electronic structure allowed by their non-trivial $\mathbb{Z}_2$ topology.~\cite{Kane:Phys.Rev.Lett.:95(2005)146802} They are therefore widely regarded ideal templates to realize the predicted exotic phenomena and applications of this topological surface state. However, real surfaces of such compounds, even if kept in ultra-high vacuum, rapidly develop a much more complex electronic structure~\cite{Bianchi:NatureCommun.:1(2010)128,King:Phys.Rev.Lett.:107(2011)096802} whose origin and properties have proved controversial.~\cite{Hsieh:Nature:460(2009)1101--1105,Bianchi:NatureCommun.:1(2010)128,King:Phys.Rev.Lett.:107(2011)096802,Bianchi:Phys.Rev.Lett.:107(2011)086802,Benia:Phys.Rev.Lett.:107(2011)177602,Wray:NaturePhys.:7(2011)32--37,Zhu:Phys.Rev.Lett.:107(2011)186405,Eremeev:arXiv:1107:3208:(2011),Menshchikova:JETPLetters:94(2011)106-111,Wray:arXiv:1105.4794:(2011),Ye:arXiv:1112.5869:(2011),Valla:Phys.Rev.Lett.:108(2012)117601,Vergniory:JETPLett.:95(2012)230--235,Chen:Proc.Nat.Acad.Sci.:109(2012)3694} Here, we demonstrate that a conceptually simple model, implementing a semiconductor-like band bending in a parameter-free tight-binding supercell calculation, can quantitatively explain the entire measured hierarchy of electronic states. In combination with circular dichroism in angle-resolved photoemission (ARPES) experiments, we further uncover a rich three-dimensional spin texture of this surface electronic system, resulting from the non-trivial topology of the bulk band structure. Moreover, our study reveals how the full surface-bulk connectivity in topological insulators is modified by quantum confinement.} 
\begin{figure*}
\begin{center}
\includegraphics[width=0.8\textwidth]{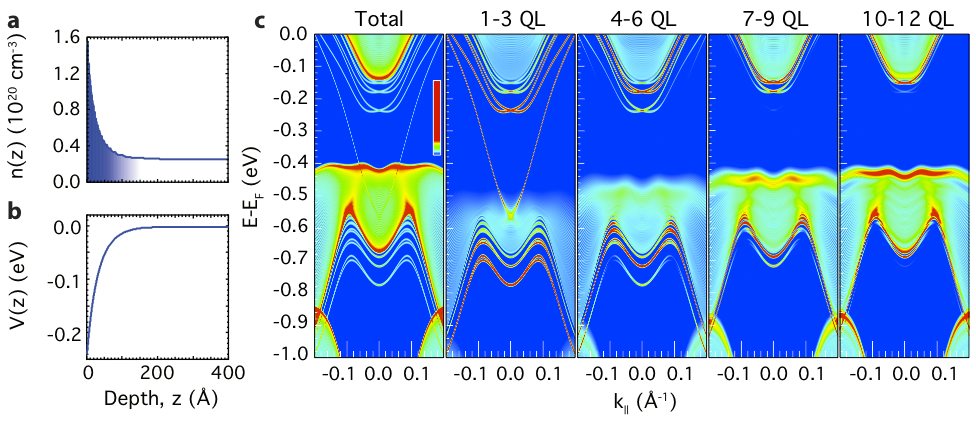}
\caption{ \label{f:confinement} {\bf Confinement of the electronic states.} (a) and (b) show the depth-dependence of the carrier density and potential profile near the surface of Bi$_2$Se$_3$, respectively. (c) Layer-projected supercell calculations reveal the varying spatial confinement of the different electronic states. In particular, the topological as well as the lower 2DEG and M-shaped quantum well states are all strongly confined within the first 3--4 quintuple layers (QLs) below the surface.}
\end{center}
\end{figure*}

Topological insulators (TIs) are an exotic state of quantum matter, guaranteed to have metallic edge or surface states due to an inverted ordering of their bulk electronic bands. The corresponding topological invariants dictate that there must be an odd number of such states intersecting the Fermi level between each pair of surface time-reversal invariant momenta (TRIM). In the most widely investigated bismuth-chalchogenide family of TIs, there is just one of these so-called topological surface states (TSSs) creating a single Dirac cone around the Brillouin zone centre.~\cite{Zhang:NaturePhys.:5(2009)438--442,Xia:NaturePhys.:5(2009)398--402,Chen:Science:325(2009)178--181,Bianchi:NatureCommun.:1(2010)128,Scanlon:AdvMat} However, exposure to even the minute amount of residual gas in an ultra-high vacuum chamber induces a drastic reconstruction of the surface electronic structure (Fig.~\ref{f:overview}). The Dirac point shifts to higher binding energies, indicating more electron-rich surfaces.~\cite{Hsieh:Nature:460(2009)1101--1105,Bianchi:NatureCommun.:1(2010)128,King:Phys.Rev.Lett.:107(2011)096802} More importantly, additional pairs of two-dimensional, almost parabolic states emerge in the vicinity of the bulk conduction band,~\cite{Bianchi:NatureCommun.:1(2010)128} which develop large Rashba-type splittings,~\cite{King:Phys.Rev.Lett.:107(2011)096802} while new ladders of M-shaped states are created in the original bulk valence bands.~\cite{Bianchi:NatureCommun.:1(2010)128,Bianchi:Phys.Rev.Lett.:107(2011)086802} Given their propensity for formation, these must be considered part of the intrinsic electronic structure of the surface of any realistic TI, for example when exposed to air or interfaced to another material. However, their existence is not predicted by idealized theoretical models of the bulk-truncated surface,~\cite{Zhang:NaturePhys.:5(2009)438--442} and has proved controversial. Recent proposals attribute their formation to a variety of electronic or structural modifications of the crystal host,~\cite{Hsieh:Nature:460(2009)1101--1105,Bianchi:NatureCommun.:1(2010)128,King:Phys.Rev.Lett.:107(2011)096802,Bianchi:Phys.Rev.Lett.:107(2011)086802,Benia:Phys.Rev.Lett.:107(2011)177602,Wray:NaturePhys.:7(2011)32--37,Zhu:Phys.Rev.Lett.:107(2011)186405,Eremeev:arXiv:1107:3208:(2011),Menshchikova:JETPLetters:94(2011)106-111,Wray:arXiv:1105.4794:(2011),Ye:arXiv:1112.5869:(2011),Valla:Phys.Rev.Lett.:108(2012)117601,Vergniory:JETPLett.:95(2012)230--235,Chen:Proc.Nat.Acad.Sci.:109(2012)3694} but to date no single model has been able to simultaneously reproduce the number and binding energies of the experimentally-observed electronic states as well as the magnitude of their measured Rashba-type splittings.

In the following, we show that this can be achieved taking account of only a single electronic phenomenon -- a near-surface electrostatic potential variation. Such intrinsic electric fields are familiar from surface space-charge regions in conventional semiconductors,~\cite{King:Phys.Rev.Lett.:101(2008)116808,Monch:Semiconductorsurfacesandinterfaces(2001)} and are also formed at interfaces where they can be modulated using applied gate voltages. Thus, our findings are not only relevant to understand the free surfaces of topological insulators, but also their use in transistor-style or other thin-film device architectures, which may ultimately provide a route towards room temperature spintronics. Here, we consider Bi$_2$Se$_3$ as a model example, but our computational approach can be freely adapted to virtually all TIs, such as other emerging Bi-chalcogenides or the half-Heusler alloys,~\cite{Lin:NatureMater.:9(2010)546--549} where we would expect a similarly striking role of band bending at free surfaces, interfaces, or in devices.

\begin{figure*}
\begin{center}
\includegraphics[width=\textwidth]{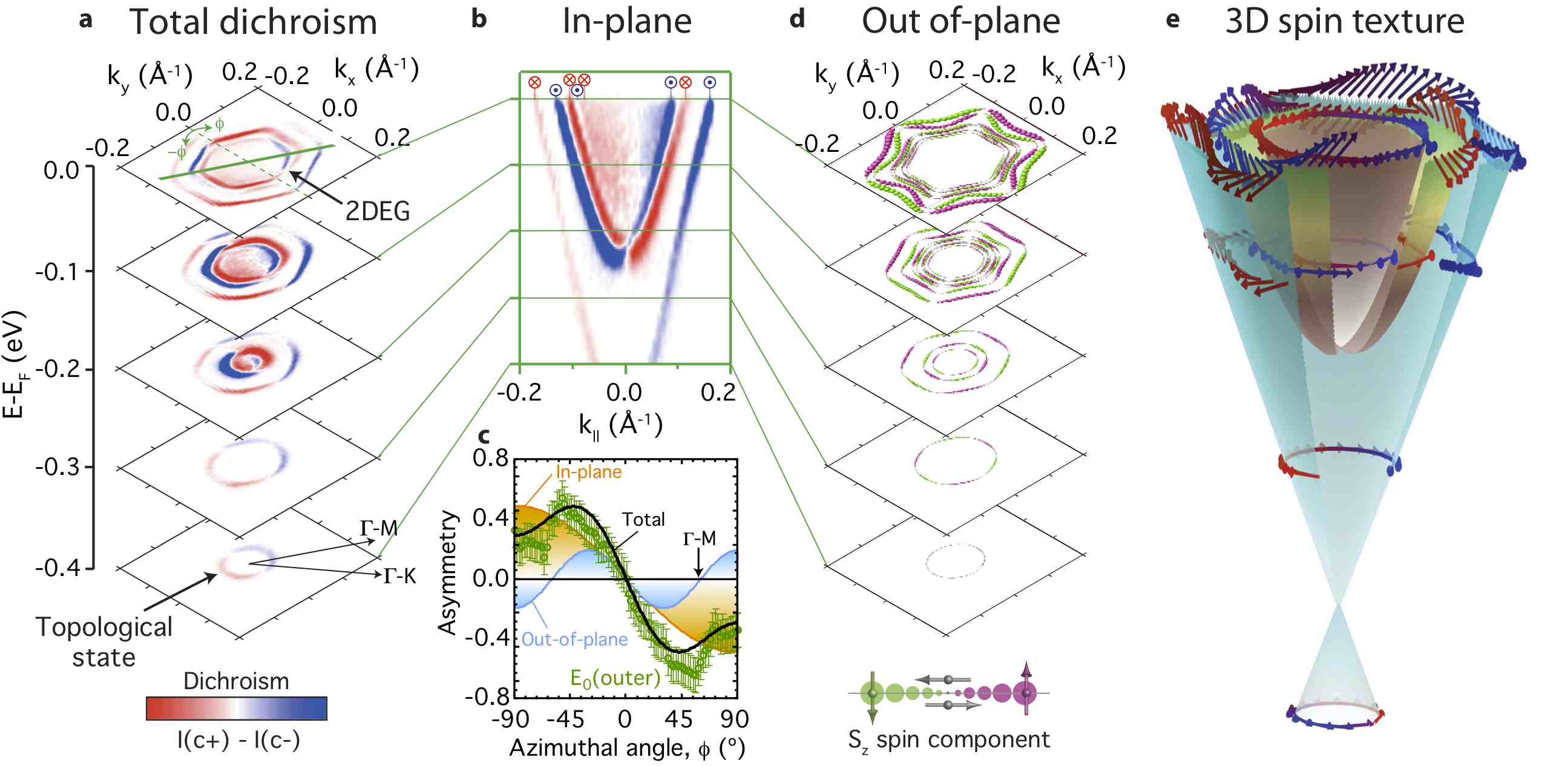}
\caption{ \label{f:spin} {\bf Three-dimensional spin texture of the topological and 2DEG states.} (a,b) Circular dichroism in constant energy contours and $E$ vs.\ $k$ dispersions measured by ARPES encode the spin texture of the topological and 2DEG states. CD along $\Gamma$-$M$ (b), containing only the in-plane component of the spin texture, revealing opposite helicity for consecutive Fermi surface sheets. (c) Quantitative analysis of the angular ($\phi$) dependence of the dichroism around the Fermi surface, shown for the outer E$_0$ 2DEG state.
A large $\sin(3\phi)$ contribution, also observed for the TSS (see Supplementary Information), reveals a significant out-of-plane spin canting of the larger Fermi surface sheets, consistent with the calculations shown in (d). In contrast, all other states largely retain the in-plane spin texture characteristic of classic Rashba systems, all the way up to the Fermi energy. Together, this leads to a rich three-dimensional spin-texture of the surface electronic structure of TIs, as summarized for the TSS and lowest ($E_0$) Rashba-split subband of the 2DEG in (e).}
\end{center}
\end{figure*}

\begin{figure*}
\begin{center}
\includegraphics[width=0.8\textwidth]{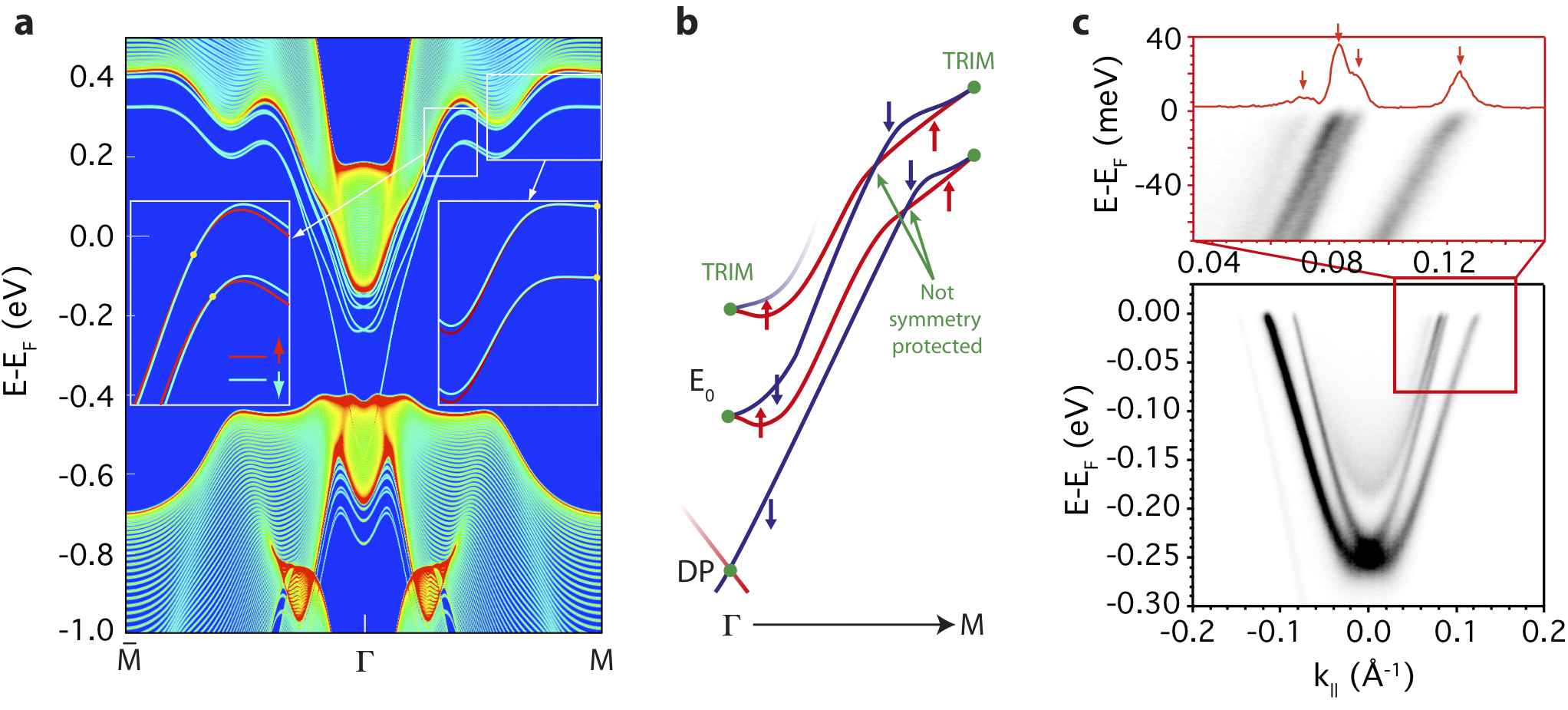}
\caption{ \label{f:connectivity} {\bf Connectivity of the surface electronic spectrum.} (a) Full electronic structure calculation along the $\Gamma$-$M$ direction. At the $\Gamma$ point, Kramer's degeneracy requires that the TSS is spin-degenerate (the Dirac point, DP) and each 2DEG state is degenerate with its spin-split pair. At the $M$ point, this degeneracy occurs between the topological state and outer branch of the 2DEG states, and then between each successive pairs of 2DEG states, as shown schematically in (b). Between these two time-reversal invariant momenta (TRIM), no fundamental symmetry dictates degeneracy of the states, and indeed these all remain non-degenerate up until at least the Fermi level, visible in high-resolution ARPES measurements (c).}
\end{center}
\end{figure*}

We implement such a band bending scenario within a 120 quintuple-layer (QL) tight-binding supercell calculation (see Methods). Without incorporating any band bending, our supercell model yields a single TSS spanning from the bulk valence to conduction bands (Supplementary Fig.~S1), in good agreement with both first-principles slab calculations~\cite{Zhang:NaturePhys.:5(2009)438--442} and our experimental measurements of the pristine surface (Supplementary Fig.~S2). Additionally including an electrostatic potential variation, we find a hierarchy of electronic states that is in excellent agreement with our experimental angle-resolved photoemission (ARPES) measurements of a doped TI surface (Fig.~\ref{f:overview}(b,c)).~\cite{fnote1} This agreement is remarkable, given that we do not adjust any parameters in our calculation. All tight-binding matrix elements are obtained from an \textit{ab-initio} calculation of the bulk electronic structure. The magnitude of the potential change is fixed by the experimentally-measured shift of the Dirac point between a pristine cleaved surface and the $n$-type doped one (see Supplementary Fig.~S2), and its functional form follows from a solution of Poisson's equation.~\cite{King:Phys.Rev.B:77(2008)125305} We stress that our approach does not require any additional modifications beyond an electrostatic potential, such as a huge increased van der Waals gap size~\cite{Eremeev:arXiv:1107:3208:(2011),Menshchikova:JETPLetters:94(2011)106-111,Ye:arXiv:1112.5869:(2011),Vergniory:JETPLett.:95(2012)230--235}, other artificial detachments of a finite number of crystal layers~\cite{Zhu:Phys.Rev.Lett.:107(2011)186405}, or the formation of new topological surface states~\cite{Wray:NaturePhys.:7(2011)32--37}.

Rather, all of the new electronic states observed experimentally arise simply due to the perturbative effect of the electrostatic potential in the system, which causes a triangular-like downward bending of the bulk electronic bands close to the surface by almost 250~meV (Fig.~\ref{f:confinement}(b)). 
Combined with the large potential step at the material/vacuum interface, this creates a near-surface quantum well for electrons. The bulk conduction band states are consequently restructured into ladders of multiple subband states.~\cite{King:Phys.Rev.Lett.:104(2010)256803,Bianchi:NatureCommun.:1(2010)128,Meevasana:NatureMater.:10(2011)114--118}  Our layer-resolved calculations (Fig.~\ref{f:confinement}) show that the lowest such subband, lying deep within the potential well, is localized in the topmost few quintuple layers (QL) of the structure: The electrons that populate this subband are free to move parallel to the surface, but are strongly confined perpendicular to it, forming a two-dimensional electron gas (2DEG). This state exhibits a strong Rashba spin-orbit splitting, whose calculated magnitude reproduces our experimental observations. Higher-lying states of the subband ladder are less strongly bound within the surface quantum well and contribute significant weight as far as 10--15~QL below the sample surface. The smaller spin splitting observed experimentally for these higher-lying states is again reproduced by our model calculations, confirming that the Rashba splitting is driven by the potential gradient of the confining electrostatic potential.~\cite{King:Phys.Rev.Lett.:107(2011)096802} Similar to the conduction band 2DEG, a ladder of subbands is also observed in the valence band near the zone center. These hole-like states become quantum confined between the surface and the upper edge of a projected bulk band gap~\cite{Zhang:NaturePhys.:5(2009)438--442,Bianchi:Phys.Rev.Lett.:107(2011)086802}, as shown schematically in Fig.~1(a). We also find a small spin splitting of these states, which is not resolved experimentally. Its location, around the top of their M-shaped dispersions, indicates a subtle interplay of spin-orbit interactions between the valence and conduction band subbands.~\cite{M.S.Bahramy:PRB84(2011)041202R}
Thus, despite its conceptual simplicity, the band bending has dramatic consequences for the surface electronic structure. Driven by an emergent role of quantum size effects, it leads to a hierarchy of electronic states in both energy and spatial extent, spanning from the two-dimensional limit of the TSS and lowest quantum well states of the conduction and valence bands to the three-dimensional continuum.

	Intriguingly, we find an anti-correlation between this electronic dimensionality, and that of the  states' spin texture. This is revealed experimentally via an asymmetry in the matrix element for photoemission when the system is excited by right and left circularly polarized light, respectively. Such circular dichroism has previously been suggested to directly~\cite{Wang:Phys.Rev.Lett.:107(2011)207602} or indirectly~\cite{Park:Phys.Rev.Lett.:108(2012)046805} probe the spin of the isolated TSS.~\cite{Wang:Phys.Rev.Lett.:107(2011)207602,Park:Phys.Rev.Lett.:108(2012)046805,Ishida:Phys.Rev.Lett.:107(2011)077601} Here, we adopt this concept to show that the full three-dimensional spin texture of the entire low-energy surface electronic structure can be extracted in a similar way. Along the $\Gamma$-$M$ mirror direction, symmetry requires that the spin must lie entirely within the surface plane.~\cite{Fu:Phys.Rev.Lett.:103(2009)266801} Our measurements of CD extracted along this direction (Fig.~\ref{f:spin}(b)) are therefore representative of the in-plane spin texture, and indicate an alternating left-right-left-right-left helicity for consecutive Fermi surface sheets of the TSS and first (E$_0$) and second (E$_1$) quantized conduction subbands. Away from this direction, dichroism of the inner E$_0$ and both E$_1$ 2DEG Fermi surface sheets vary sinusoidally with angle, $\phi$, around the Fermi surface (shown in Supplementary Information). This reveals a model Rashba-like~\cite{Bychkov:JETPLett.:39(1984)78} helical spin texture of these states, with the spin polarised entirely in the surface plane at all energies. However, for the more spatially-localized TSS and outermost 2DEG states, the dichroism additionally develops a strong $\sin(3\phi$) component which is maximal along $\Gamma$-$K$ and zero along $\Gamma$-$M$ (Fig.~\ref{f:spin}(c) and Supplementary Information). This reflects a strong out-of-plane spin polarization close to the Fermi level, which is correlated with hexagonal warping of the electronic states,~\cite{Fu:Phys.Rev.Lett.:103(2009)266801} and again switches sign between neighbouring states. Thus, not only does the band bending promote the creation of additional Fermi surface sheets, it also leads to complex band- and binding-energy dependent three-dimensional spin-textures of the resulting electronic system, as summarized in Fig.~\ref{f:spin}(e). Together, this will cause significant complications for the interpretation of spin-dependent transport experiments, and inter-band processes will weaken the topological protection against backscattering commonly assumed for the TSS of these compounds. Nonetheless, we note that our resolution limited linewidths of $\lesssim0.006$~\AA$^{-1}$ ($\lesssim0.13^\circ$) indicate that the intrinsic scattering remains relatively weak for all Fermi surface sheets.

The non-trivial spin texture of these states also raises questions over their interplay. For any system with TR symmetry, the electronic states must be spin-degenerate at TRIM, even though a breaking of inversion symmetry by the confining potential and surface allows a lifting of the spin degeneracy at arbitrary $k$-points. The hexagonal surface Brillouin zone of topological insulators such as Bi$_2$Se$_3$ contains TRIM $k$-points at the zone-centre ($\Gamma$ point) and at the side-centre ($M$) point. Fig.~\ref{f:connectivity}(a) shows the electronic structure calculated along the entire $\Gamma$-$M$ direction. At the $\Gamma$ point, each Rashba-split pair of 2DEG states clearly become degenerate. For the TSS, this initially appears to be complicated by the presence of bulk states, with the Dirac point looking to be buried deep in the bulk valence band. However, locally, the TSS is still situated within a projected band gap of the (bent) bulk bands, and so remains a well defined surface state (unlike the 2DEG states which are located within, and indeed derived from, the projected bulk-like bands). Indeed, our layer-projected calculations (Fig.~\ref{f:confinement}) reveal that the Kramer's degeneracy of the TSS is preserved in the presence of the band bending potential, contrary to previous suggestions of a band gap opening in the surface spectrum at the Dirac point.~\cite{Wray:NaturePhys.:7(2011)32--37,Wray:arXiv:1105.4794:(2011)} 

At the $M$-point, our calculations reveal that the TSS connects to the outer branch of the lower 2DEG state. Assuming a spin-down state for the TSS, the pairing 2DEG state has no other choice but to be spin-up. Its spin-down partner must then become connected with the outer branch of the second subband to satisfy TR symmetry for these states, thereby dictating the overall up-down-up-down-up spin configuration discussed above. Fundamentally, this pair-exchange follow from the non-trivial $\mathbb{Z}_2$ topology of the bulk band structure. This requires that the TSS must connect from the bulk valence to conduction bands. Now, however, the original bulk bands have become quantized into two-dimensional subband states in the vicinity of the surface, and our calculations reveal that it is to these subbands which the topological state connects. We note that this requires the outer and inner branches of the 2DEG states to diverge from each other at large $k$-vectors, in stark  contrast to conventional Rashba systems~\cite{Ishizaka:Nat.Mater.:10(2011)521} in which the spin-split bands merge again at zone-edge TRIM.

Hexagonal warping of the electronic states can additionally induce accidental degeneracies along high-symmetry lines,~\cite{Bahramy:Nat.Commun:3(2012)679,Basak:PRB84:(2011)121401(R)} as seen in Fig.~\ref{f:connectivity}. However, as these are not protected by any fundamental symmetry, introduction of perturbations in the system can shift, or even completely remove, such accidental degeneracies. Experimentally, the inner branch of the lower 2DEG state and outer branch of the upper 2DEG state can be seen to approach each other at low binding energies (Fig.~\ref{f:connectivity}(c)), consistent with their proximity to the accidental crossing predicted in our calculations. However, contrary to previous reports~\cite{Wray:NaturePhys.:7(2011)32--37,Wray:arXiv:1105.4794:(2011)}, our high-resolution measurements reveal that the occupied bands remain nondegenerate and exhibit rather parallel dispersion close to the Fermi level, which could be indicative of an avoided crossing. Understanding whether interactions not included in our theoretical treatment play a role in avoiding the accidental degeneracies will be crucial to unravel surface-bulk coupling in these materials, and the role of interactions in limiting mobilities of spin-polarized surface transport in topological insulators.

{\small \subsection*{Methods}
 {\it Theory:} Relativistic electronic structure calculations were carried out within the context of density functional theory using the Perdew-Burke-Ernzerhof exchange-correlation functional~\cite{Perdew:PRL:77(1996)3865} and the augmented plane wave plus atomic orbitals (APW-LO) method as implemented in the WIEN2K program.~\cite{wien2k}  For all the atoms, the muffin-tin radius $R_{MT}$ was set to 2.5 Bohr and the maximum modulus of reciprocal vectors $K_{max}$ was chosen such that $R_{MT}K_{max}=7.0$. The primitive cell was considered to be hexagonal (space group: $R\bar{3}m$) with lattice parameters and atomic positions taken from experiment.~\cite{Vicente:InrogChem:38(1999)2131} The corresponding BZ was sampled using a $10\times10\times3$ $k$-mesh. To simulate the effect of band bending, a 120 QL tight-binding supercell Hamiltonian was constructed by downfolding the APW-LO Hamiltonian using maximally localized Wannier functions (MLWF's).~\cite{souza,mostofi,kunes} We chose valence $p$ orbitals of Bi and Se as the projection centers of MLWF's. The  bending potential was obtained by solving the coupled Poisson-Schr\"odinger equation,~\cite{King:Phys.Rev.B:77(2008)125305} assuming  a static dielectric constant of 70.~\cite{King:Phys.Rev.Lett.:107(2011)096802} The resulting potential was then  added to the onsite terms of the  tight-binding sucpercell Hamiltonain. The (QL-projected) surface band structure was eventually obtained by diagonalizing the supcercell Hamiltonian (and projecting the resulting eigenstates onto the MLWF's of respective surface QL's).

{\it Experiment:} Bi$_2$Se$_3$ crystals were prepared by reacting high purity elements (5N) of Bi and Se in sealed, evacuated quartz tubes at 850$^\circ$C for 2 days, followed by cooling at 2-3$^\circ$C/h to 650$^\circ$C. The crystals were annealed at this temperature for 7 days before quenching to room temperature. ARPES measurements were performed at 10~K using circularly-polarized light with photon energies of 20--30~eV and a Scienta R4000 hemispherical analyser at the SIS beamline of the Swiss Light Source. Surface doping was achieved by depositing potassium on freshly-cleaved samples from a properly outgassed SAES K getter source. }

{\small \subsection*{Acknowledgements}
We are grateful for support from the UK EPSRC, the ERC, and the Japan Society for the Promotion of Science (JSPS) through the `Funding Program for World-Leading Innovative R\&D on Science and Technology (FIRST Program)', initiated by the council for Science and Technology Policy (CSTP). }

\bibliographystyle{naturemag}

\end{bibunit}

\

\begin{bibunit}
\begin{center}{\large \bf Supplementary Information}\end{center}
\renewcommand{\thefigure}{S\arabic{figure}}
\setcounter{figure}{0}   
\renewcommand{\theequation}{S\arabic{equation}}  
\renewcommand{\thesection}{SI \arabic{section}}

\section{Electron doping of the surface}

\begin{figure}
\begin{center}
\includegraphics[width=0.9\columnwidth]{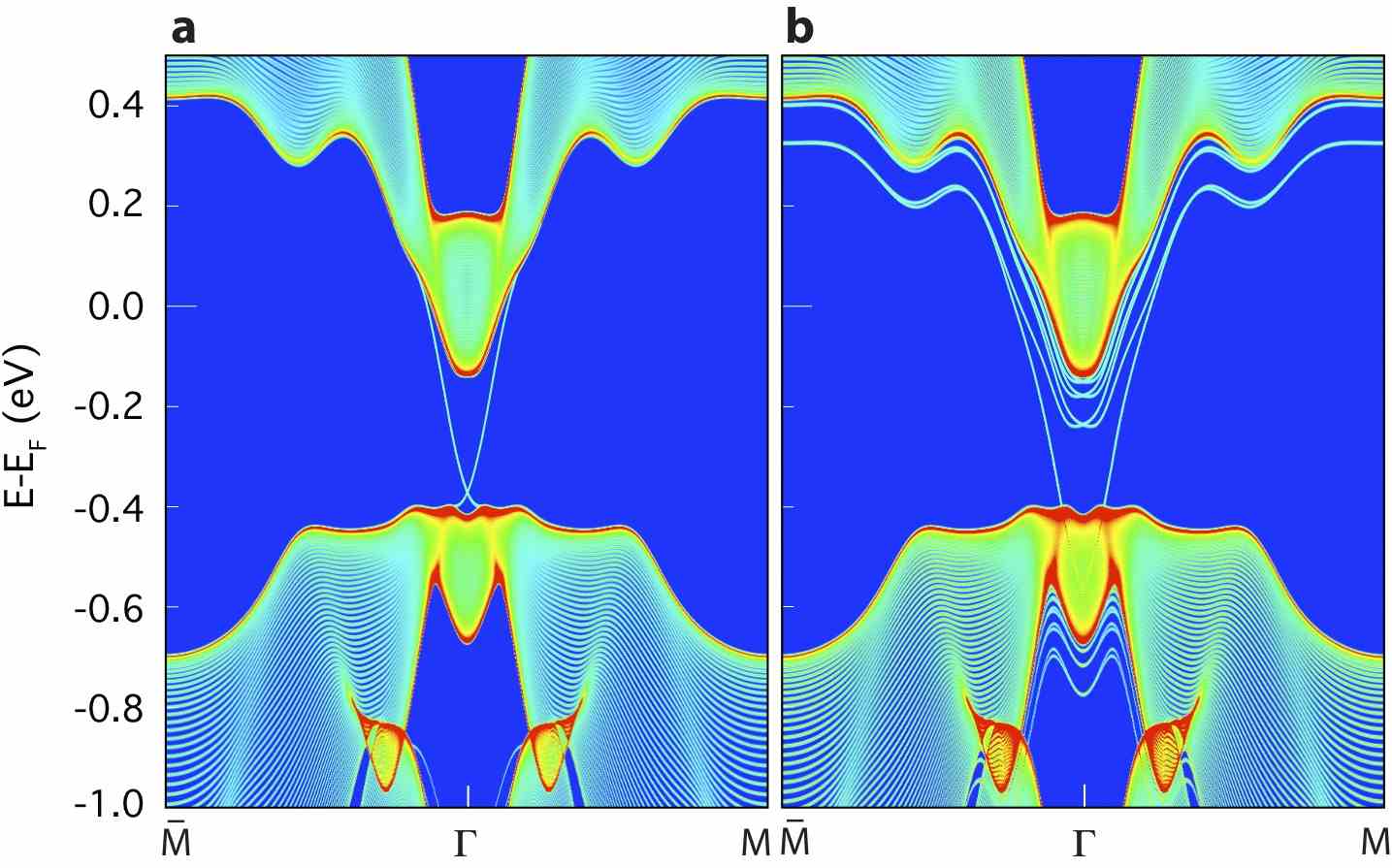}
\caption{ \label{f:bending_effect} Projected band structure from a 120QL {\it ab-initio}-derived tight-binding model of the ideal Bi$_2$Se$_3$ structure (a) without and (b) including an additional near-surface band bending.}
\end{center}
\end{figure}

The surface projection of our {\it ab-initio}-derived tight-binding model for the ideal bulk-truncated surface of Bi$_2$Se$_3$ is shown in Fig.~\ref{f:bending_effect}(a). Besides the projected bulk bands, the topological surface state gives rise to a clear Dirac-like feature spanning from the bulk valence to conduction bands, in good agreement with previous idealized first-principles models of the surface electronic structure of topological insulators.~\cite{Zhang:NaturePhys.:5(2009)438--442} A very similar electronic structure is also seen in our experimental measurements of the pristine surface of Bi$_2$Se$_3$, performed immediately after cleaving the sample in ultra-high vacuum. Additionally imposing a near-surface electrostatic potential in our model ({\it i.e.,} a downward band bending resulting from surface doping) leads to a drastic reconstruction of the electronic structure, as discussed in the main text and shown for comparison in Fig.~\ref{f:bending_effect}(b).

\begin{figure}
\begin{center}
\includegraphics[width=\columnwidth]{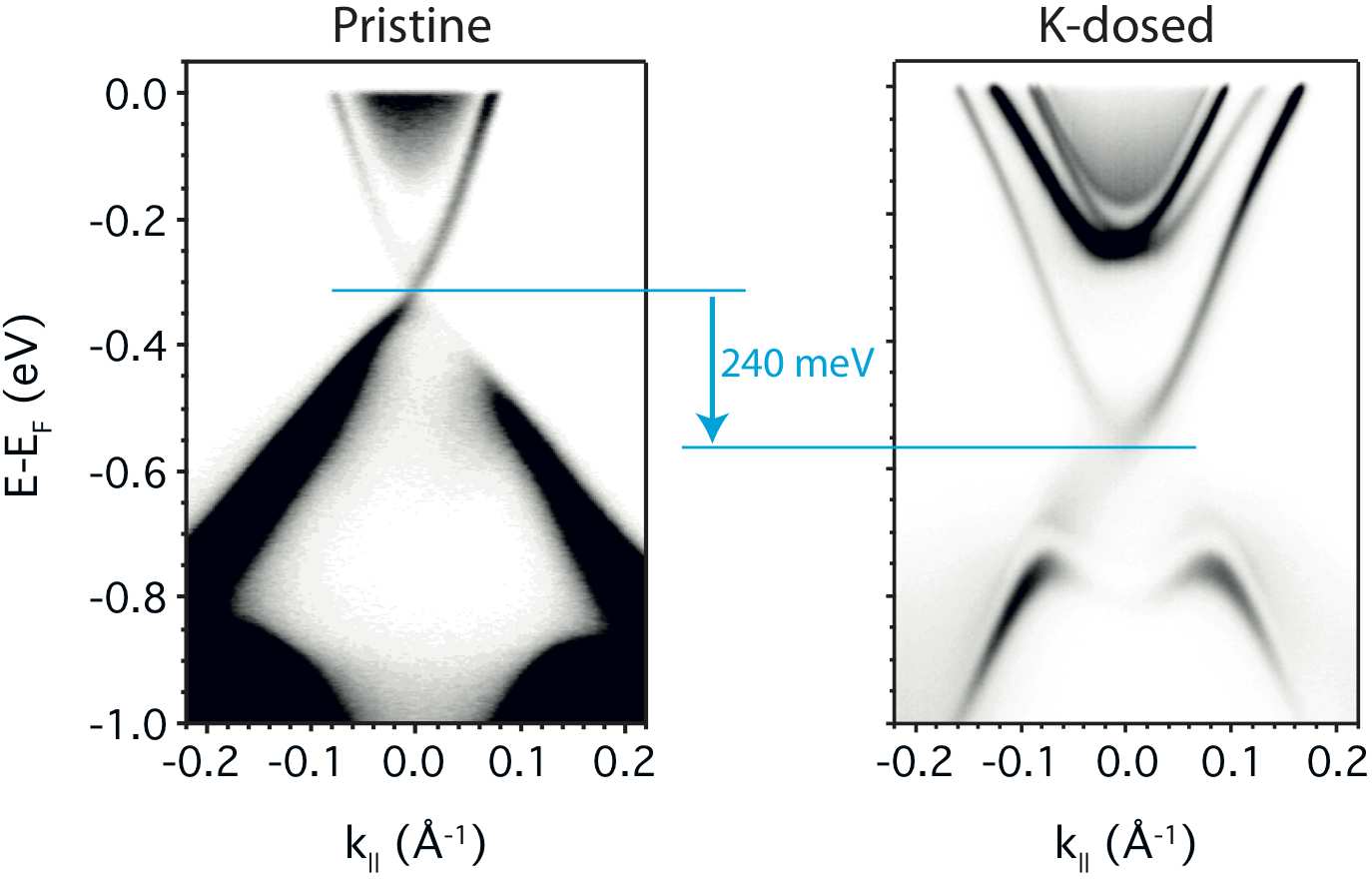}
\caption{ \label{f:doping_comp} ARPES spectra taken from the pristine (left) and K-dosed (right) surface. The Dirac point shifts downwards by 240~meV, accompanied by the quantization of the bulk conduction and valence band states.}
\end{center}
\end{figure}

Surface doping via adsorption of residual gas from the vacuum onto the surface of Bi$_2$Se$_3$, and related compounds, is already sufficient to drive time-dependent changes of the electronic structure, which ultimately leads to such surface quantum well formation.~\cite{Bianchi:NatureCommun.:1(2010)128,King:Phys.Rev.Lett.:107(2011)096802} However, creation of such band bending can be drastically accelerated by deliberately adsorbing minute quantities of a range of molecules or atoms on the surface.~\cite{King:Phys.Rev.Lett.:107(2011)096802,Bianchi:Phys.Rev.Lett.:107(2011)086802,Benia:Phys.Rev.Lett.:107(2011)177602,Wray:NaturePhys.:7(2011)32--37,Zhu:Phys.Rev.Lett.:107(2011)186405} Here, we employ alkali metal dosing, depositing $\ll0.1$~ML (from XPS analysis) of potassium on the freshly-cleaved sample surface from a properly outgassed SAES K getter source. This causes the spectral changes shown in Fig.~\ref{f:doping_comp}, including a shift of the Dirac point to higher binding energies by 240~meV, and the emergence of Rashba-split conduction band subbands and M-shaped valence band subbands, as also seen in our theoretical model (Fig.~\ref{f:bending_effect}(b)) and discussed in detail in the main text.

\section{Quantitative analysis of Circular Dichroism}
\begin{figure*}
\begin{center}
\includegraphics[width=0.73\textwidth]{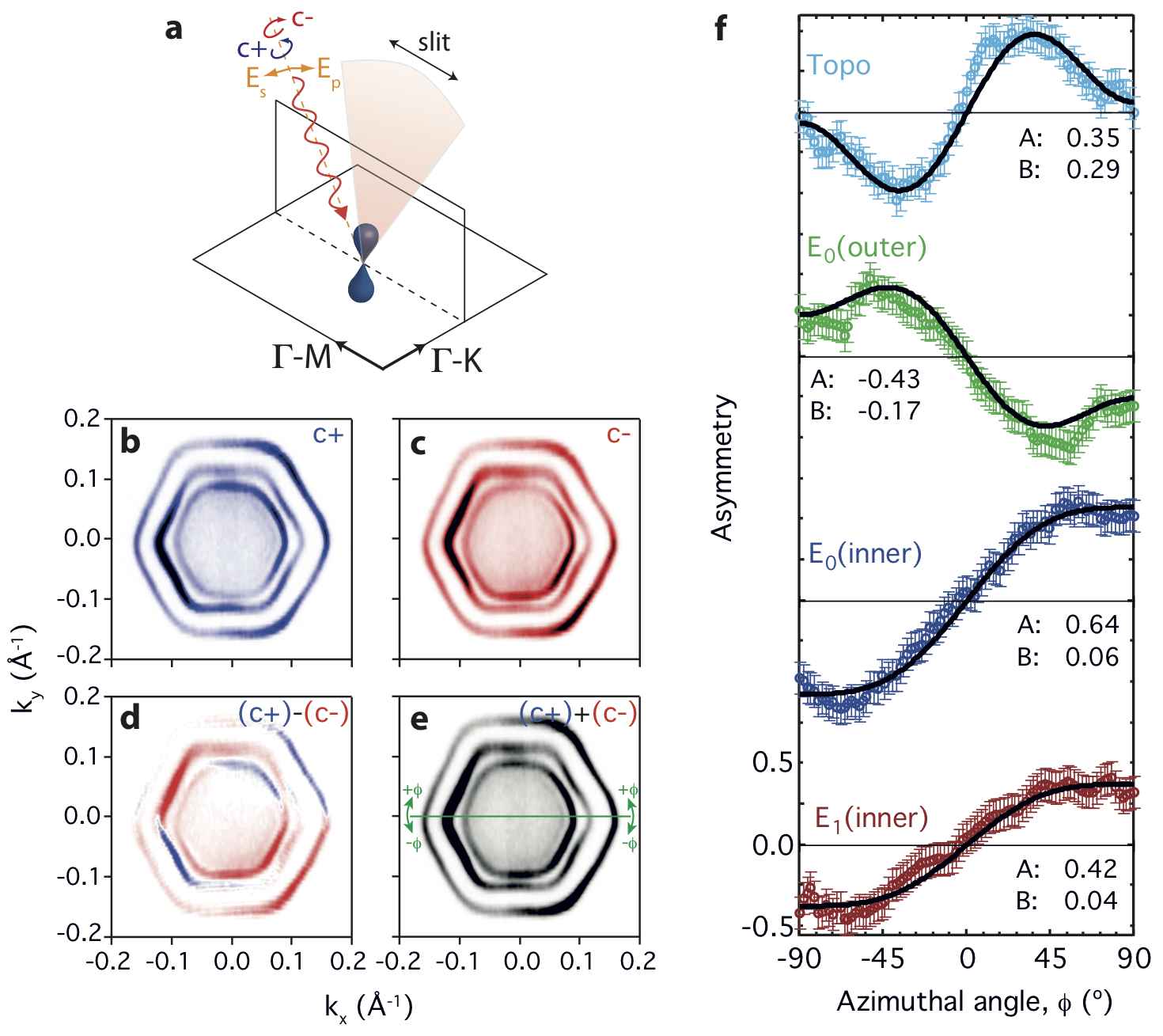}
\caption{ \label{f:FS_asym} Circular dichroism measurements of the Fermi surface, performed using the experimental geometry shown in (a). Fermi surface map, measured using (b) right-hand and (c) left-hand circularly polarized light, respectively. These give the dichroism and sum shown in (d) and (e), respectively. (f) Measured asymmetry (defined in the text) of the topological, Rashba-split inner and outer branch of the lower 2DEG and innermost branch of the upper 2DEG states. These are vertically offset for clarity, and fit by $A\sin(\phi)+B\sin(3\phi)$ (solid black lines), with the coefficients given in the figure.}
\end{center}
\end{figure*}

The experimental geometry for the circular dichroism measurements is shown in Fig.~\ref{f:FS_asym}(a). The analyser entrance slit is oriented along the $\Gamma$-$M$ direction, and the full electronic structure is measured by tilting the sample around this axis. A clear intensity difference can be observed between the Fermi surface measured in this way using right and left hand circularly polarized light, respectively (Fig.~\ref{f:FS_asym}(b-d)). Such circular dichroism (CD) has recently been shown to encode the spin-texture of the topological surface state (TSS) in topological insulators (TIs).~\cite{Wang:Phys.Rev.Lett.:107(2011)207602,Park:Phys.Rev.Lett.:108(2012)046805,Ishida:Phys.Rev.Lett.:107(2011)077601,Jung:Phys.Rev.B:84(2011)245435} We use this method to additionally extract the spin texture of the 2DEG states, as sumarized in the main manuscript. We note that, for the relative changes in CD between the bands that we study here we can neglect final-state effects, as transitions from all of these closely-spaced initial states will go into the same (broad) final state.

\begin{figure*}
\begin{center}
\includegraphics[width=0.95\textwidth]{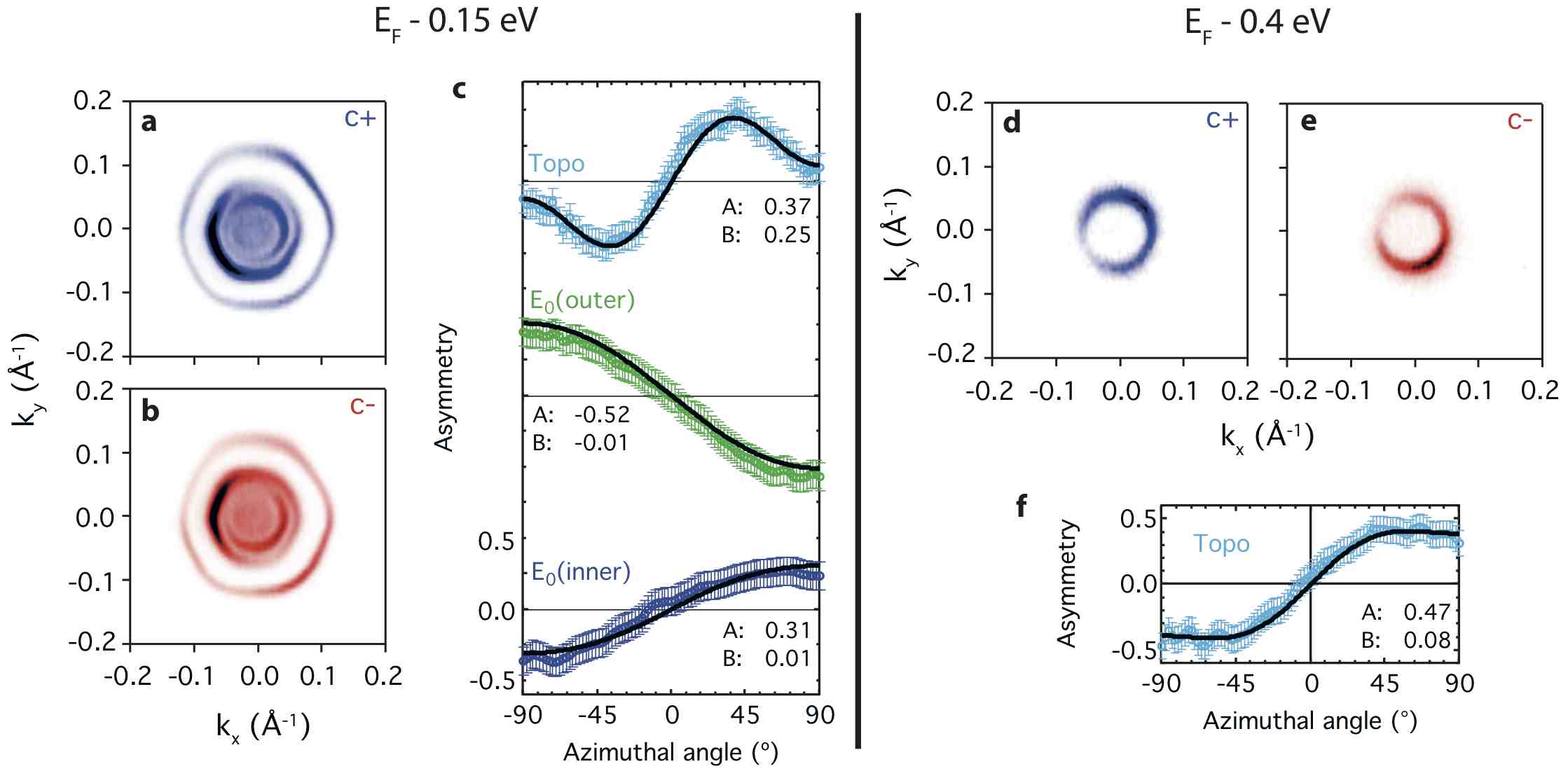}
\caption{ \label{f:Eb_asym} Constant energy contours from 0.15~eV below the Fermi level performed using (a) right-hand and (b) left-hand circularly polarized light, respectively. (c) Extracted asymmetry and model fits for the topological and lower 2DEG states, vertically offset for clarity. (d-f) Equivalent measurements for 0.4~eV below the Fermi level, where only the topological state is present.}
\end{center}
\end{figure*}

To enable quantitative analysis, we define an asymmetry parameter
\[A(\phi)=\frac{I^{c+}(\phi)-I^{c-}(\phi)}{I^{c+}(\phi)+I^{c-}(\phi)},\]
where $I^{c\pm}(\phi)$ is the normalized intensity of photoemission measured using right or left circularly polarized light, respectively, at an azimuthal angle $\phi$ around a Fermi surface contour (defined in Fig.~\ref{f:FS_asym}(e)), with a node in the CD always found for $\phi=0$. Such asymmetry measurements are shown at the Fermi level in Fig.~\ref{f:FS_asym}(f) and at binding energies of 0.15~eV and 0.4~eV in Fig.~\ref{f:Eb_asym}(c) and (f), respectively. These are fit by the function $A\sin(\phi)+B\sin(3\phi)$, with the $A,B$ coefficients shown in the figures. CD varying with a simple $\sin(\phi)$ dependence around the Fermi surface arises from the in-plane winding of the spin around the Fermi surface, while the $\sin(3\phi)$ component results from coupling to the out-of-plane spin component.~\cite{Wang:Phys.Rev.Lett.:107(2011)207602,Jung:Phys.Rev.B:84(2011)245435,Fu:Phys.Rev.Lett.:103(2009)266801} The ratio of the $A$ and $B$ components in the fit therefore gives a good measure of the degree of out-of-plane spin polarization. We note that this is the minimal model consistent with the $C_{3v}$ symmetry of the surface~\cite{Fu:Phys.Rev.Lett.:103(2009)266801}. Negligible improvement in the fits was found when including possible higher-order terms (such as a $\sin(6\phi)$ component~\cite{Jung:Phys.Rev.B:84(2011)245435,Basak:Phys.Rev.B:84(2011)121401}), so we neglect these in the following.

The asymmetries around the Fermi surface are shown in Fig.~\ref{f:FS_asym}(f). The asymmetry changes sign upon moving inwards from the outermost Fermi surface sheet of the topological state to the subsequent Fermi surfaces of the lowest Rashba-split subband ($E_0$) of the 2DEG. This is entirely consistent with the results of spin-resolved photoemission for these states.~\cite{King:Phys.Rev.Lett.:107(2011)096802} While it is difficult to resolve the outer branch of the second Rashba-split subband ($E_1$), its spin-split pair is visible as the innermost Fermi surface sheet at positive $k_x$ values in Fig.~\ref{f:FS_asym}(b--e). The asymmetry of this state has the same sign as the inner branch of the lower subband state (Fig.~\ref{f:FS_asym}(f)), indicating an overall up-down-up-down-up ordering of the spin for consecutive Fermi surface pockets.

The asymmetry of the inner branch of both the lower and upper ($E_0$ and $E_1$) subbands is well described by a single sinusoid, with negligible $\sin(3\phi)$ component in the fits. The spin therefore lies almost entirely within the surface plane for these near-circular Fermi surfaces, as expected for a simple model Rashba system.~\cite{Bychkov:JETPLett.:39(1984)78} On the other hand, the asymmetries of the outer branch of the lower ($E_0$) subband and the TSS both contain a substantial component which varies as $\sin(3\phi)$, indicating a significant canting of the spin out of the surface plane. This is entirely consistent with the strong hexagonal warping of their Fermi surfaces.~\cite{Fu:Phys.Rev.Lett.:103(2009)266801,Frantzeskakis:Phys.Rev.B:84(2011)155453,Souma:Phys.Rev.Lett.:106(2011)216803,Xu:Science:332(2011)560--564}

At energies close to the band bottom of the 2DEG states, such an out-of-plane spin polarization is still present for the TSS (Fig.~\ref{f:Eb_asym}(c)). Now, however, the spin lies within the surface plane not only for the inner but also the outer branch of the lowest 2DEG subband. At higher binding energies still, well below the band bottom of the 2DEG states and approaching the Dirac point, the ideal in-plane helical spin texture is recovered also for the topological state (Fig.~\ref{f:Eb_asym}(f)).

Therefore, not only the direction of the spin winding, but also its effective dimensionality, has a distinct variation with binding energy for each of the multiple electronic bands that make up the complex surface electronic structure of these materials.

\bibliographystyle{naturemag}

\begin{thebibliography}{10}
\bibitem{fnote1} The effective mass is slightly overestimated in our model, which stems from a small overestimation of this quantity in our bulk LDA calculations.
\end{thebibliography}

\end{bibunit}

\end{document}